\documentclass[aps,preprint]{revtex4}

\usepackage{amssymb}
\usepackage{amsmath}
\usepackage{epsfig}
\usepackage{epstopdf}
\usepackage{bm}
\usepackage{graphicx,epsfig}
\usepackage{mathrsfs}
\usepackage{dcolumn}
\usepackage{color}
\usepackage{natbib}
\usepackage{CJK}
\usepackage{booktabs}
\usepackage{diagbox}
\def\be{\begin{equation}}
\def\ee{\end{equation}}
\def\bea{\begin{eqnarray}}
\def\eea{\end{eqnarray}}

\allowdisplaybreaks[2]
\begin{document}
\title{Optical neural network architecture for deep learning  with the temporal synthetic dimension}
\author{Bo Peng$^{1,\dag}$, Shuo Yan$^{1,\dag}$, Dali Cheng$^2$, Danying Yu$^{1}$, Zhanwei Liu$^{1}$, Vladislav V. Yakovlev$^3$,  Luqi Yuan$^{1,*}$, and Xianfeng Chen$^{1,4,5}$}
\affiliation{$^1$State Key Laboratory of Advanced Optical Communication Systems and Networks, School of Physics and Astronomy, Shanghai Jiao Tong University, Shanghai 200240, China \\
$^2$Ginzton Laboratory and Department of Electrical Engineering, Stanford University,\\
 Stanford,CA,49305, USA\\
$^3$Texas A\&M University, College Station, Texas 77843, USA\\
$^4$Shanghai Research Center for Quantum Sciences, Shanghai 201315, China\\
$^5$Collaborative Innovation Center of Light Manipulation and
Applications,
Shandong Normal University, Jinan 250358, China\\
$^\dag$These authors contribute equally to this work.
$^*$Corresponding author: yuanluqi@sjtu.edu.cn\\}

\date{\today }

\begin{abstract}
The physical concept of synthetic dimensions has recently been
introduced into optics. The fundamental physics and applications
are not yet fully understood, and this report explores an approach
to optical neural networks using synthetic dimension in time
domain, by theoretically proposing to utilize a single resonator
network, where the arrival times of optical pulses are
interconnected to construct a  temporal synthetic dimension. The
set of pulses in each roundtrip therefore provides the sites in
each layer in the optical neural network, and can be linearly
transformed with splitters and delay lines, including the phase
modulators, when pulses circulate inside the network. Such linear
transformation can be arbitrarily controlled by applied modulation
phases, which serve as the building block of the neural network
together with a nonlinear component for pulses. We validate the
functionality of the proposed optical neural network  for the deep
learning purpose with examples handwritten digit recognition and
optical pulse train distribution classification problems. This
proof of principle computational work explores the new concept of
developing a photonics-based machine learning in a single ring
network using synthetic dimensions, which allows flexibility and
easiness of reconfiguration with  complex functionality in
achieving desired optical tasks.

\end{abstract}

\maketitle

\section{Introduction}

Optical neural networks (ONN) are under extensive studies recently
with an ultimate goal of achieving machining learning in a
photonic system
\cite{rosenbluth09,tait14,shenNP17,tait17,linS18,yingOL18,feldmann19,zuoO19,hamerly19,khoram19,zhang19newa,zhang21newa}.
Recent advancements have revealed that ONN exhibits important
computation capability with photonic tools
\cite{nahmias20,wetzstein20,bogaerts20,xuN21,feldmann21} and
training optical fields for some specific optimization purposes
\cite{jiangNRM20}. On the other hand, realizations of ONN on
different platforms  also attract great interest from theoretical
and computational perspectives. For example, training ONN through
in situ back propagation \cite{hughes18,zhouPR20} and quantum ONN
can conduct the non-classical tasks \cite{stenbrecher20}.  In
addition, the recurrent neural network \cite{I. Goodfellow 17, K.
Yao11, G. Dorffner13, M. Husken12, J. T. Connor14}, as an
important machine learning model, has been studied with the
optical-based technologies \cite{by Hugeyyy}. Nevertheless, it has
been found that most of ONN designs depend on the number of
photonic devices in each layer as well as the total layer number,
which makes an ONN system require $N^2$ photonic devices with
tunable externally controlled components and makes its practical
implementation rather complex and lacks the freedom and options
for further reconfiguration and miniaturization
\cite{nahmias20,wetzstein20,bogaerts20}. It is therefore important
to investigate alternative photonic ONN design architectures,
which can potentially offer enough freedom towards arbitrary
functionality. Thus, it is essential to explore novel physical
principles, and the approach based on synthetic dimensions offers
an intriguing opportunity to overcome some of the existing
challenges and limitations.

Synthetic dimension is a rapidly-arising concept in photonics
which facilitates utilization of different degrees of freedom of
light to simplify experimental arrangements and get the most out
of those \cite{yuanoptica18,ozawaNRP19,Topological photonics,
Chinese Optics Letters, APL photonics 6}. Recently, it has been
suggested that  ONN with  synthetic dimensions can potentially
provide simpler design of the ONN to achieve a complicated
functionality \cite{pankov19,buddhiraju20,linArxiv20}. However,
the proper implementation of those appeared to be challenging. In
this report, we investigate the time-multiplexed architecture
using temporal information
\cite{regensburger11,regensburger12,wimmer13,marandi14,wimmer17,chenPRL18}
that has been demonstrated a highly promising way for optical
computations such as coherent Ising machines \cite{marandi14},
photonic reservoir computing \cite{23newlarger17}, and ONN with
synthetic nonlinear lattices \cite{arxiv9-8Aus}.

In this work, we introduce and validate through computational
experiments a new paradigm to achieve the optical neural network
in a single  resonator network, with the temporal synthetic
dimension constructed by connecting different temporal positions
of pulses with pairs of delay lines. Different from pioneering
works in Refs. \cite{pankov19,arxiv9-8Aus} that propose ONN with
synthetic lattices in coupled rings, the proposed approach here
offers an alternative solution to the ONN problem in a single
ring. The optical resonator network with reconfigurable couplings
between different arrival times (i.e., temporal positions) of
optical pulses supports time-multiplexed lattice \cite{marandi14}
and creates the temporal synthetic dimension. With controllable
splitters and phase modulators used to build desired connections
between pulses, we show the way of constructing multiple layers of
ONN in a single resonator (see Fig. \ref{scheme}(a)). A nonlinear
operation is used to perform complex modulations which are being
controlled by external signals with the aid of a computer. As
validations for the deep-learning functionality, we perform the
training of the proposed platform for ONN with the training data
set of MNIST handwritten digit database with appropriate noises
considered \cite{zzzmnist}.  The striking feature of our ONN is
that it needs only one resonator but gives arbitrary size of
layers in the network, which makes our system unlimited in total
layer (roundtrip) number with high reconfigurability. Moreover,
this single resonator network is capable of conducting arbitrary
optical tasks, after performing the proper training. For example,
we conduct a pulse train classification problem, which recognizes
different distributions of pulse trains.  Our work hence points
out a concept for realizing the ONN with  synthetic dimensions,
which is highly scalable and therefore gives the extra freedom for
further simplification of the setup with possible reconfiguration.

\section{Model}

\begin{figure}[!ht]
\centering
\includegraphics[width=1\linewidth]{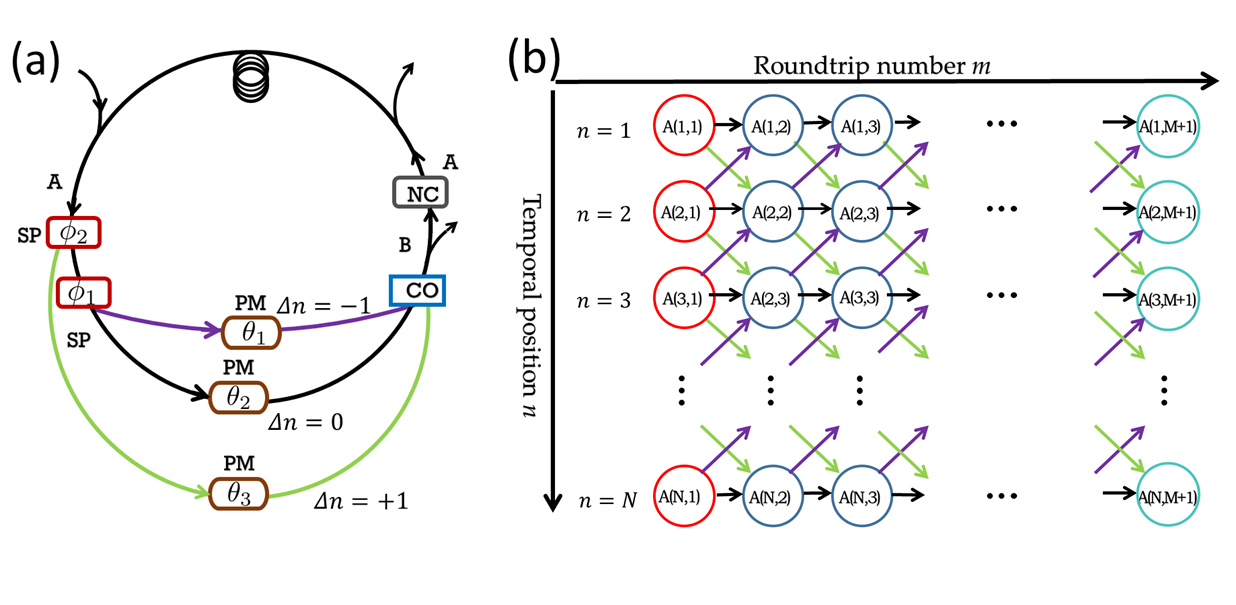}
\caption{(a) The schematic of the single resonator network with
two delay lines in purple and green respectively. CO: Combiner,
SP: Splitter, PM: Phase modulator, NC: Nonlinear component. $A$
denotes to the field amplitude while $B$ denotes output amplitude
defined in Eq. \ref{linear}. (b) The connectivity of the synthetic
photonic lattice along the temporal dimension ($n$-axis)
implemented in (a)  for pulses evolving after roundtrips ($m$). A
number of $N$ pulses in each roundtrip (shown in circles) is
considered and the pulses evolve for $M$ roundtrips in total,
which therefore constructs the ONN with $M$ layers and $N$ neurons
sites in each layer. Green, black, and purple arrows correspond to
different optical branches of delay lines in (a).  \label{scheme}}
\end{figure}

We start considering a resonator composed of the main cavity loop
of the waveguide [see Fig. \ref{scheme}(a)]. By neglecting the
group velocity dispersion of the waveguide, we assume  there are
$N$ optical pulses  simultaneously propagating inside the loop,
and every two nearby pulses is temporally separated by a fixed
time $\Delta t$. Each pulse is labelled by its temporal position
$t_n$ (or arrival time, with $t_{n+1}-t_n = \Delta t$)
\cite{marandi14}, and we  use $n=1,...,N$ to denote each pulse at
different temporal positions.

To construct the temporal synthetic dimension, we add a pair of
delay lines, which are connected with the main  loop through
splitters and couplers. Each splitter is controlled by parameter
$\phi_{1(2)}$, which determines that a portion of the pulse with
the amplitude $\cos \phi_{1(2)}$ remains in the main loop while
the rest of the pulse with the amplitude $i\sin \phi_{1(2)}$ gets
into the delay line \cite{regensburger11,regensburger12}. Lengths
of  delay lines are carefully designed. For the pulse at the
temporal position $n$ propagating through the shorter delay line,
it combines into the main loop at a time $\Delta t$ ahead of its
original arrival time $t_n$ and contributes to the pulse at the
time $t_{n-1} = t_n - \Delta t$, i.e., $\Delta n = -1$. On the
other hand, for the pulse propagating through the longer delay
line, it combines into the main loop at a time $\Delta t$ behind
$t_n$ and contributes to the pulse at  $t_{n+1} = t_n + \Delta t$,
i.e., $\Delta n = +1$. Such a design constructs the temporal
synthetic dimension [see Fig. \ref{scheme}(b)], where the $n$-th
pulse during the $m$-th roundtrip with the amplitude $A(n,m)$ (in
a unit of a reference amplitude $A_0$) is connected to its nearest
neighbor sites in the temporal synthetic lattice after each
roundtrip. The boundary of this lattice can be created by further
introducing the intracavity intensity modulator to suppress
unwanted pulses in the main loop \cite{leefmans}.

We place phase modulators inside the main loop as well as two
delay lines. Each phase modulator is controlled by external
voltage and  adds a modulation phase $\theta_i$ ($i=1,2,3$) for
the pulse propagating through it  \cite{marandi14,leefmans}.
Moreover, we use the complex modulator as the nonlinear component,
which can convert the input pulse to an output pulse with a
complex nonlinear function. In such ONN, parameters $\phi_i$ and
$\theta_i$ can be precisely controlled at any time, meaning that
one can manipulate $\phi_i$ and $\theta_i$ for each pulse $n$ at
each roundtrip number $m$.

In this temporal synthetic  lattice, the propagation process of
pulses in each single roundtrip can compose the linear
transformation, described by \cite{regensburger11,regensburger12}
\begin{equation*}
    B(n,m) = A(n,m)\cos\phi_1(n,m)\cos\phi_2(n,m)e^{i\theta_2(n,m)}
\end{equation*}
\begin{equation*}
  -iA(n+1,m)\sin\phi_2(n+1,m)e^{i\theta_3(n+1,m)}
\end{equation*}
\begin{equation}
  -iA(n-1,m)\cos\phi_2(n-1,m)\sin\phi_1(n-1,m)e^{i\theta_1(n-1,m)}, \label{linear}
\end{equation}
where $B(n,m)$ denotes output amplitudes for the set of pulses
after the linear transformation. A very small portion of pulses
are dropped out and collected by detectors, which are stored in
the computer for the further analysis. The pulses then pass the
nonlinear component where we use a formula has similar formula as
a saturable absorber \cite{baoNR11,chengIEEE14} but with
amplitudes, so a complex nonlinear operation is performed
\begin{equation}
    2B(n,m)(1-T_{n,m})/A_0 =\mathrm{ln} (T_{n,m}), \label{nonlinear1}
\end{equation}
\begin{equation}
        A(n,m+1) = B(n,m)T_{n,m},  \label{nonlinear2}
\end{equation}
For a given input pulse $B(n,m)$, the nonlinear coefficient
$T_{n,m}$ can be calculated in the computer with Eq.
(\ref{nonlinear1}), and then appropriate external signal is
applied to the complex modulator \cite{nonlinearnew3} so the
output pulse after the nonlinear component follows Eq.
(\ref{nonlinear2}), which turns out to be the input pulse
$A(n,m+1)$ for the next layer (the next roundtrip).  We find that
this particular choice of the complex nonlinear function works
extremely well, compared to regular real nonlinear activation
functions such as sigmoid function or  hyperbolic tangent
function.

\begin{figure}[!ht]
\centering
\includegraphics[width=1\linewidth]{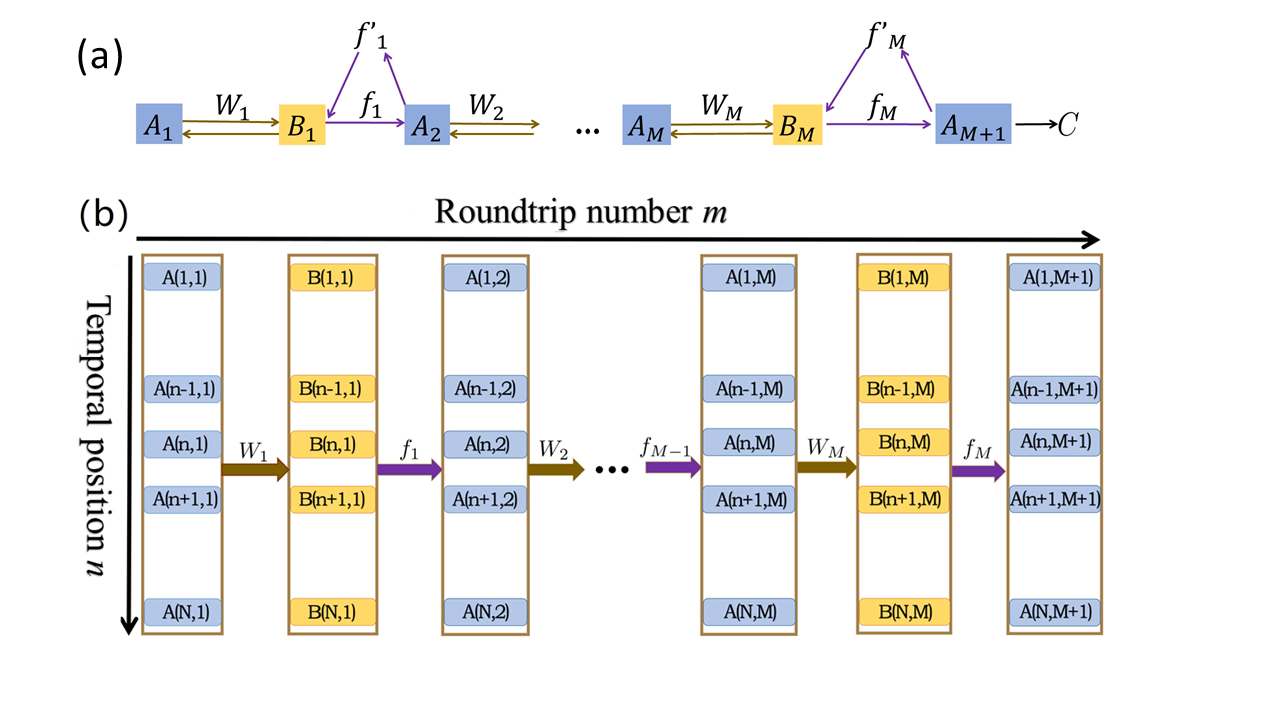}
\caption{(a) Schematic of the architecture of an optical neural
network. $A_1$ is the vector of pulses imported in the first layer
when training starts. $A_m$: vector of the output pulses after the
$(m-1)$-th roundtrip (layer), which is also the input vector for
the $m$-th roundtrip (layer); $W_m$: matrix for the linear
transformation during the $m$-th roundtrip (layer); $B_m$: vector
of pulses after the linear transformation during the $m$-th
roundtrip (layer); $f_m$: nonlinear activation operation; $f_m'$:
derivative of $f_m$ during back propagation. $C$ is the cost
function for the output signal. (b) Illustration of the signal
flow through roundtrips (layers) in the resonator in Fig.
\ref{scheme}(a). \label{ONNnetwork}}
\end{figure}

Fig. \ref{ONNnetwork} summarizes the forward transmissions with
linear transformations and nonlinear operations on pulses.
Theoretically, the total number of layers, $M$ as well as the
total pulse number $N$,  can be arbitrary. In Fig.
\ref{ONNnetwork}, we use $W_m$ to define the linear transformation
in Eq. (\ref{linear}) and $f_m$ to define the nonlinear operation
in Eqs. (\ref{nonlinear1}) and (\ref{nonlinear2}) for the $m$-th
roundtrip. Hence the forward transmission at each layer $m$
follows $B_m=W_m A_m$ and $A_{m+1} =f_mB_m$, where $A_m$ and $B_m$
are vectors of $A(n,m)$ and $B(n,m)$, respectively. Pulse
information $A(n,m+1)$ ($B(n,m)$) after (before) the nonlinear
operation at the $n$-th temporal position during the $m$-th
roundtrip is collected by dropping a small portion of pulses out
of the resonator network into detectors. Such information of $A_m$
and $B_m$ is stored in the computer for further backward
propagation in training the ONN.

Once the forward propagation is finished after $M$ roundtrips in
the optical resonator network, the backward propagation can be
performed in the computer following the standard procedure to
correct control parameters  \cite{hughes18,bengio09}, which is
briefly summarized here. The backward propagation equations read
\cite{hughes18,bengio09}:
\begin{equation}
     \tilde{B}_m=B_m+f'_m(A_{m+1}-\tilde{A}_{m+1}),
     \label{backward1}
\end{equation}
\begin{equation}
        \tilde{A}_m=W_m^{T}\tilde{B}_,
        \label{backward2}
\end{equation}
$\tilde A_m$ and $\tilde B_m$ are vectors at the $m$-th layer,
calculated through the back propagation from the stored
information of $A_{m+1}$ and $B_m$. Here $f'_m$ is the derivative
of the nonlinear operation at the $m$-th layer in Eq.
(\ref{backward1}), $W_m^{T}$ is the inverse of $W_m$, and $\tilde
A_{M+1}$ is the target vector $A_{\mathrm{target}}$, which is the
expected output vector of the training set. The cost function
after the $m$-th layer can therefore be calculated as:
\begin{equation}
        {C_m = \frac{1}{2N} \sum_{i=1}^N|A(i,m+1)- \tilde{A}(i,m+1)|^2}.\\
        \label{cost}
\end{equation}
Throughout the backward propagation, optical controlling
parameters $\phi_1(n,m)$, $\phi_2(n,m)$, $\theta_1(n,m)$,
$\theta_2(n,m)$, and $\theta_3(n,m)$ can be trained by calculating
the derivative of $C_m$ with respect to these parameters, i.e.,
\begin{equation}
     \frac{\partial  C_m}{\partial \phi_{1,2}(n,m)}=[(A_{m+1}-\tilde{A}_{m+1})]^T\bigodot f'_m \cdot \frac{\partial W^T}{\partial \phi_{1,2}(n,m)}\cdot A_m,
     \label{daoshu1}
\end{equation}
\begin{equation}
    \frac{\partial  C_m}{\partial \theta_{1,2,3}(n,m)}=[(A_{m+1}-\tilde{A}_{m+1})]^T\bigodot f'_m \cdot \frac{\partial W^T}{\partial \theta_{1,2,3}(n,m)}\cdot A_m,
    \label{daoshu2}
\end{equation}
$\bigodot$ is the vector multiplication, with $\textbf{c} =
\textbf{a} \bigodot \textbf{b}$ defined as $c_n = a_n b_n$. We can
obtain the corrections of parameters as \cite{bengio09}:
\begin{equation}
        \Delta\phi_{1,2}(n,m)=-a\frac{\partial C_m}{\partial \phi_{1,2}(n,m)},
         \label{correct1}
\end{equation}
\begin{equation}
        \Delta\theta_{1,2,3}(n,m)=-a\frac{\partial C_m}{\partial \theta_{1,2,3}(n,m)},
         \label{correct2}
\end{equation}
where $a$ is learning rate for this training. Then
$\phi_{1,2}(n,m)$ becomes $\phi_{1,2}(n,m)+\Delta \phi_{1,2}(n,m)$
and $\theta_{1,2,3}(n,m)$ becomes $\theta_{1,2,3}(n,m)+\Delta
\theta_{1,2,3}(n,m)$. Following the backward propagation procedure
summarized above, the parameters for controlling the forward
propagation of each pulse at the $n$-th temporal position for the
$m$-th roundtrip are updated backwardly from the $M$-th layer to
the $1$-st layer.

Having the entire procedure in hand,  one can train ONN with a
training set of data to prepare the ONN ready for  doing the
designed all-optical computation with optical pulses in this
single resonator network.

\section{Results}

\subsection{Handwritten digit recognition}

\begin{figure}[!ht]
\centering
\includegraphics[width=1\linewidth]{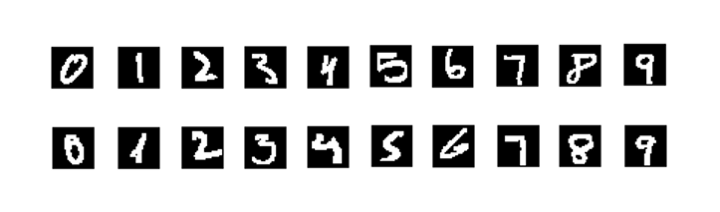}
\caption{ Typical visualization legends from the MNIST dataset.}
\label{Typicallegend}
\end{figure}

To show the validity and reliability of our proposed ONN, we
consider a MNIST handwritten digit recognition problem as commonly
used for ONN \cite{zzzmnist}, with noises included. The MNIST data
set is chosen from the classic data set in the field of machine
learning. It consists of 60000 training samples and 10000 test
samples. Each sample is a 28 * 28 pixel grayscale handwritten
digital picture, representing a number from 0 to 9. Some typical
visualization legends are given in  Fig. \ref{Typicallegend}.

\begin{figure}[!ht]
\centering
\includegraphics[width=1\linewidth]{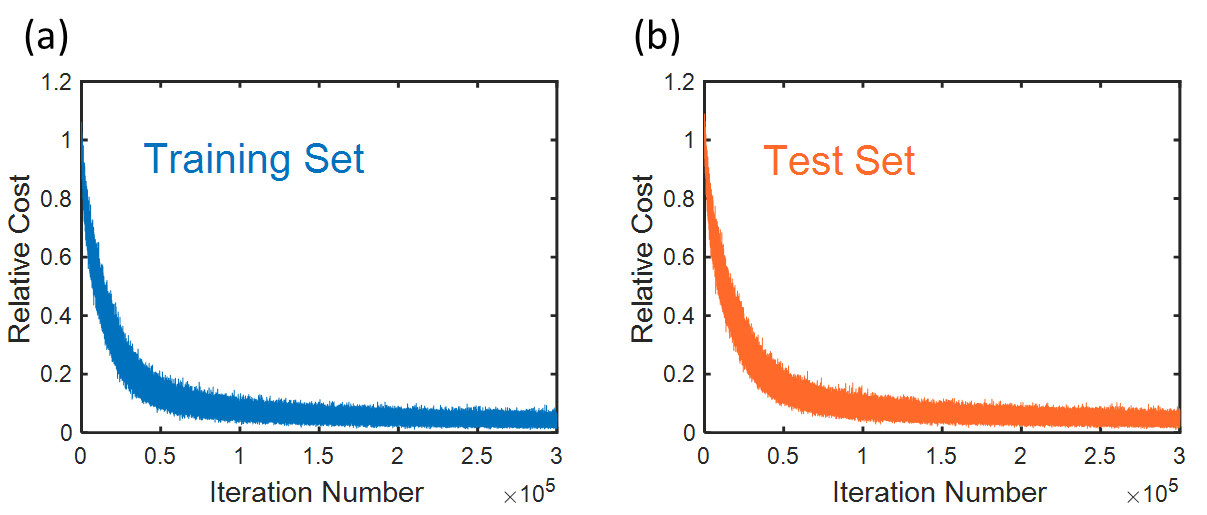}
\caption{ Relative cost functions defined in Eq. (\ref{cost})
versus the training iteration number during the training process
or  (a) training set, and (b) test set, respectively, for the Hand
written recognition problem. } \label{costfigure}
\end{figure}

In simulations, we use $49$ pulses ($N=49$) and $45$ roundtrips
($M=45$), with learning rate $a=0.001$. For the simplicity
purpose, we pre-process the original MNIST handwritten digit
database \cite{zzzmnist}, where each input data supposes to have
an array of $784$ elements, with the maximum pooling twice
\cite{pooling}, so the input data can be mapped on $49$ input
pulses in our ONN architecture. Moreover, after the final
roundtrip in the single resonator network, we add another full
connection layer between collected signals from $49$ pulses and
$10$ additional output sites, which shall be assisted by the
computer. In this full connection layer on the computer, $a=0.02$
and we use the sigmoid nonlinear function as the activation
function. In Fig. \ref{costfigure}, we plot the normalized cost
function, defined in Eq. (\ref{cost}), for training set and test
set versus the computation iteration number, respectively. Such
cost function based on the mean square error has been used in the
literature for classification problems
\cite{newRaudys,newLehtokangas,newSaleem,newSebastiani}. Both cost
functions decreases as the iteration number increases. Therefore,
the ONN in the  temporal synthetic dimension works fine for the
handwritten digit recognition problem. We emphasize  that the
pre-processing and the additional full connection layer make this
model less competitive with previous ONN \cite{previous
onn2,linS18,feldmann19}, this simulation is only for the purpose
of demonstrating the validity of our ONN and the stability with
certain noises. To this end, in the test set, we add random noises
on the $49$ input pulses with their amplitudes multiplied by
$1+R\cdot \delta/2$, where $R\in (-0.5,0.5)$ is a random number
and $\delta$ denotes  amplitude of noises, where we choose
$\delta=0,2\%,4\%,6\%,8\%,10\%$, respectively. $60000$ sets of
training data and $10000$ sets of test data are used for
simulations. After training, noises with $\delta$ are appended
into the ONN to do the test. We list errors of prediction in Table
\ref{table1}. One can see that the error of prediction in our ONN
architecture is $21.1\%$ if there is no noise in input pulses from
the test set. However, when we add noises into the system, the
error increases up to $29.7\%$ for $\delta=10\%$. Small noises may
be tolerated in this proposed ONN architecture. However large
noises could affect the performance of the system, which might
need further improve in the future. Although the effects of noises
in our proposed ONN architecture are difficult to be compared with
those in other ONN systems due to the very different design
associated to synthetic dimensions, typical experiments with
time-multiplexed architecture can be done with small noises
\cite{marandi14}.

\begin{table}
  \centering
    \caption{Errors of prediction for handwritten digit recognition with different noises. }
    \setlength{\tabcolsep}{3.5mm}
  \begin{tabular}{c|c| c| c| c |c| c}
      \hline
      $\delta$ in test set (\%) & 0 & 2 & 4 & 6 & 8 & 10\\
      \hline
      error of prediction (\%) & 21.1 & 24.6 & 25.8 & 27.1 & 28.0 & 29.7 \\
      \hline
  \end{tabular}
\label{table1}
\end{table}

\subsection{Optical pulse train distribution classification}

We have demonstrated the  validity of the proposed ONN. One of the
key importance of this proposal is to provide a possible trained
optical network to act a certain photonic functionality
intelligently. As a  simple proof-in-principle verification, we
perform a home-made optical pulse train distribution
classification problem.

Our goal is to train an optical neural network to recognize five
different profiles of optical pulse trains composed by $101$
pulses, where  shapes of five profiles are chosen as sinusoidal
functions  as $\sin(k \pi t_i/T)$ for  pulse at  temporal position
$t_i$, with $T=100\Delta t$ and $k=1,2,3,4,5$ labelling five
profiles respectively. For both training and test procedures, each
pulse is interrupted with  noises. $30000$ training sets and
$5000$ test sets are constructed in  simulations. Similar noise is
used so pulse is modified in amplitude by
$1+R_{1(2)}\delta_{1(2)}$, where $\delta_{1(2)}$ is  amplitude of
noises in the training (test) sets and $R_{1(2)}\in (-0.5,0.5)$ is
a random number. The choice of $\delta_{1(2)}$ is listed in Table
\ref{table2}.

During simulations, $101$ pulses ($N=101$) and $31$ roundtrips
($M=31$) are chosen for the ONN, and after the final roundtrip,
another full connection layer between $101$ pulses and 5 output
sites is used for predictions. For the training procedure, the
learning rate $a$ is $0.001$ for $\delta_1=0$, $0.0017$ for
$\delta_1 =2\%$, $0.021$ for $\delta_1=4\%$, and $0.011$ for
$\delta_1=6\%$. The choice of $\delta_1=0$ results in the
invalidation due to lack of data type (all training sets having
same labels are identical), while the noise amplitude
$\delta_1=6\%$ induces the complexity caused by high volatility
and instability of our data. Therefore, for these two training
procedures, the test result has relatively high errors of
prediction. Nevertheless, one can see from Table \ref{table2}
that, for noise amplitudes $\delta_1=2\%$ and $4\%$ in the
training procedures, the errors of prediction in the test
procedures show relatively good results (error of prediction
$\lesssim 30\%$), even for high noise amplitudes $\delta_2=14\%$
in the test procedures.

\begin{table}
  \centering
    \caption{Errors of prediction for optical pulse train distribution classification problems. }
    \setlength{\tabcolsep}{7mm}
  \begin{tabular}  {c|c|c|c|c}
      \hline
      \diagbox[width=50mm, height=20mm] {$\delta_2$ (\%)}{ $\delta_1$  (\%) } & 0 & 2 & 4 & 6 \\
      \hline
      0 & 1.1 & 16.6 & 24.4 & 29.8  \\
      \hline
      2 & 43.7 & 18.7 & 26.6 & 30.8  \\
      \hline
      4 & 68.9 & 19.9 & 26.7 & 30.9  \\
      \hline
      14 & 75.4 & 30.7 & 29.4 & 32.4  \\
      \hline
  \end{tabular}
  \label{table2}
\end{table}

The training process with zero noise of pulses in the training set
is invalidated due to the monotonicity of the data set. However,
in the case of low noise amplitudes of pulses in the training set,
our ONN system shows a relatively stable prediction for optical
pulse train distribution classifications. Furthermore, as another
important feature, one can see that, for a larger noise amplitude
$\delta_1$ in the training set (for example, compare
$\delta_1=2\%$ and $4\%$), although it gives larger errors for
smaller noise amplitudes $\delta_2$ in the test set, one can
obtain a smaller error (such as $30.7\%$ and $29.4\%$) for the
relative large noise $\delta_2=14\%$. The example therefore shows
the capability of our proposed ONN architecture in performing
direct optical processing.

\section{Discussion and Summary}

The proposed platform is experimentally feasible with the
state-of-the-art photonic technology. The fiber ring resonator
with kilometer-long roundtrip length can be constructed with
hundreds of temporal separated pulses circulating inside the
resonator \cite{marandi14,leefmans}. In particular, Ref.
\cite{leefmans} shows the capability for constructing a 64
time-multiplexed optical resonant sites with pulses produced by an
input 1550 nm mode-locked laser, separated by 4 ns, which points
out an excellent possible experimental platform for realizing our
theoretical proposal. Moreover, this proposal for achieving the
temporal synthetic dimension can also be realized in a resonator
with the free-space optics \cite{chenPRL18}. In both setups, delay
lines (channels) are used to create the nearest-neighbor couplings
along the temporal synthetic dimension. Moreover, appropriate
delay lines (channels) can also connect pulses at time separations
with double, triple, and/or high-order $\Delta t$, i.e., providing
the long-range couplings. It therefore holds the possibility for
generating more than three connectivities between sites in two
layers in Fig. \ref{scheme}(b), which might be possible to further
increase the accuracy of the ONN. These delay lines may induce
small errors, but as one sees in Table \ref{table1} and Table
\ref{table2}, the synthetic ONN can tolerate small noises. The
current nonlinear function in the proposal is performed in the
computer. However, it is possible to consider nonlinear component
operated by amplitude and phase modulations
\cite{nonlinearnew3,nonlinearnew1,nonlinearnew2} or other
nonlinear components \cite{arxiv9-8Aus,IEEE-William}, which can
perform alternative different complex nonlinear functions in
optics. One notices that, in the proposed approach, the back
propagation in the training process is conducted with a computer
and then obtained optimal parameters are transferred to the
physical system. Such ex-situ training might bring extra errors,
but is currently a reasonable strategy utilized in recent
experiments for demonstrating ONN functionality
\cite{shenNP17,linS18,nahmias20,feldmann19}. Ref. \cite{hughes18}
suggests a possible way to realize in-situ backward propagation in
optical systems, which may greatly improve speed in ONN. The
inclusion of such in-situ backward propagation in our proposed ONN
could be of interest for the future study.

In summary, we propose a novel paradigm to achieve the ONN in a
single resonator network. The proposed approach is based on a
physical concept of the temporal synthetic dimension.  As the
proof of principle, we study the MNIST handwritten digit
recognition problem to verify the validation of the deep learning
functionality of our proposed ONN. Furthermore, we  demonstrate
the possibility of photonic intelligent features, by showing the
performance of a home-made optical pulse train distribution
classification problem. Our proposed ONN in the temporal synthetic
dimension uses the trade-off between time and space complexity,
and therefore does not have the advantages in energy and speed.
However, the key achievement here is that we propose an
alternative model with relatively high flexibility, which can be
re-configurable and scalable on the number of sites (pulses) in
each layers as well as the number of layers (roundtrips) for each
computation. Distinguished from other relevant works
\cite{pankov19,arxiv9-8Aus}, our proposal focuses one resonator
supporting temporal synthetic dimension and shows the opportunity
for constructing a flexible ONN that is capable for various
optical tasks once getting trained. The construction in Fig.
\ref{scheme}(b) can be easily linked to architectures of
conventional neural networks with long-range connectivities added
via additional delay lines, which can be further generalized to a
recurrent neural network  \cite{I. Goodfellow 17, K. Yao11, G.
Dorffner13, M. Husken12, J. T. Connor14}. Furthermore, one can
also prepare the set of pulses with the single-photon state
instead \cite{chenPRL18}, which might makes our proposal with the
temporal synthetic dimension being possible for constructing the
quantum neural network in the future study. Our work therefore
shows the opportunity for constructing a flexible ONN in a single
resonator, which points to a broad range of potential applications
from all-optical computation to intelligent optical information
processing \cite{chensegev} and biomedical imaging
\cite{newSierra,newShirshin}.

\begin{acknowledgments}
The research was supported by National Natural Science Foundation
of China (12122407, 11974245, and 12192252), Shanghai Municipal
Science and Technology Major Project (2019SHZDZX01-ZX06). V. V. Y.
acknowledges partial funding from NSF (DBI-1455671, ECCS-1509268,
CMMI-1826078), AFOSR (FA9550-15-1-0517, FA9550-18-1-0141,
FA9550-20-1-0366, FA9550-20-1-0367), DOD Army Medical Research
(W81XWH2010777), NIH (1R01GM127696-01, 1R21GM142107-01), and the
Cancer Prevention and Research Institute of Texas (RP180588). L.Y.
thanks the sponsorship from Yangyang Development Fund and the
support from the Program for Professor of Special Appointment
(Eastern Scholar) at Shanghai Institutions of Higher Learning.
\end{acknowledgments}


\begin{thebibliography}{99}
\bibitem{rosenbluth09} D. Rosenbluth, K. Kravtsov, M. P. Fok, and P. R. Prucnal, ``A high performance photonic pulse processing device,'' Optics Express \textbf{17}, 22767--22772 (2009).

\bibitem{tait14} A. N. Tait, M. A. Nahmias, B. J. Shastri, and P. R. Prucnal, ``Broadcast and weight: an integrated network for scalable photonic spike processing,'' Journal of Lightwave Technology \textbf{32}, 4029--4041 (2014).

\bibitem{shenNP17} Y. Shen, N. C. Harris, S. Skirlo, M. Prabhu, T. Baehr-Jones, M. Hochberg, X. Sun, S. Zhao, H. Larochelle, D. Englund, and M. Solja\v{c}i\'{c}, ``Deep learning with coherent nanophotonic circuits,'' Nature Photonics \textbf{11}, 441--446 (2017).

\bibitem{tait17} A. N. Tait, T. F. de Lima, E. Zhou, A. X. Wu, M. A. Nahmias, B. J. Shastri, and P. R. Prucnal, ``Neuromorphic photonic networks using silicon photonic weight banks,'' Scientific Reports \textbf{7}, 7430 (2017).

\bibitem{linS18} X. Lin, Y. Rivenson, N. T. Yardimci, M. Veli, Y. Luo, M. Jarrahi, and A. Ozcan, ``All-optical machine learning using diffractive deep neural networks,'' Science \textbf{361}, 1004--1008 (2018).

\bibitem{yingOL18} Z. Ying, Z. Wang, Z. Zhao, S. Dhar, D. Z. Pan, R. Soref, and R. T. Chen, ``Silicon microdisk-based full adders for optical computing,'' Optics Letters \textbf{43}, 983--986 (2018).

\bibitem{feldmann19} J. Feldmann, N. Youngblood, C. D. Wright, H. Bhaskaran,
and W. H. P. Pernice, ``All-optical spiking neurosynaptic networks
with self-learning capabilities,'' Nature \textbf{569}, 208--214
(2019).

\bibitem{zuoO19} Y. Zuo, B. Li, Y. Zhao, Y. Jiang, Y.-C. Chen, P. Chen, G.-B. Jo, J. Liu, and S. Du, ``All-optical neural network with nonlinear activation functions,'' Optica \textbf{6}, 1132--1137 (2019).

\bibitem{hamerly19} R. Hamerly, L. Bernstein, A. Sludds, M. Solja\v{c}i\'{c}, and D. Englund, ``Large-scale optical neural networks based on photoelectric multiplication,'' Physical Review X \textbf{9}, 021032 (2019).

\bibitem{khoram19} E. Khoram, A. Chen, D. Liu, L. Ying, Q. Wang, M. Yuan, and Z. Yu, ``Nanophotonic media for artificial neural inference,''  Photonics Research \textbf{7}, 823--827 (2019).


\bibitem{zhang19newa} T. Zhang, J. Wang, Y. Dan, Y. Lanqiu, J. Dai, X. Han, X. Sun, and K. Xu,  ``Efficient training and design of photonic neural network through neuroevolution,'' Optics Express \textbf{27}, 37150--37163 (2019).

\bibitem{zhang21newa} H. Zhang, J. Thompson, M. Gu, D. Jiang, H. Cai, P. Y. Liu, Y. Shi, Y. Zhang, M. F. Karim, G. Q. Lo, X. Luo, B. Dong, L. C. Kwek, and A. Q. Liu, ``Efficient On-Chip Training of Optical Neural Networks Using Genetic Algorithm,'' ACS Photonis \textbf{8}, 1662--1672 (2021).

\bibitem{wetzstein20} G. Wetzstein, A. Ozcan, S. Gigan, S. Fan, D. Englund, M. Solja\v{c}i\'{c}, C. Denz, D. A. B. Miller, and D. Psaltis, ``Inference in artificial intelligence with deep optics and photonics,'' Nature \textbf{588}, 39--47 (2020).

\bibitem{nahmias20} M. A. Nahmias, T. F. de Lima, A. N. Tait, H.-T. Peng, B. J. Shastri, and P. R. Prucnal, ``Photonic multiply-accumulate operations for neural networks,'' IEEE Journal of Selected Topics in Quantum Electronics \textbf{26}, 1--18 (2020).


\bibitem{bogaerts20} W. Bogaerts, D. P\'{e}rez, J. Capmany, D. A. B. Miller, J. Poon, D. Englund, F. Morichetti, and A. Melloni, ``Programmable photonic circuits,'' Nature \textbf{586}, 207--216 (2020).

\bibitem{xuN21} X. Xu, M. Tan, B. Corcoran, J. Wu, A. Boes, T. G. Nguyen, S. T. Chu, B. E. Little, D. G. Hicks, R. Morandotti, A. Mitchell, and D. J. Moss, ``11 TOPS photonic convolutional accelerator for optical neural networks,'' Nature \textbf{589}, 44--51 (2021).

\bibitem{feldmann21} J. Feldmann, N. Youngblood, M. Karpov, H. Gehring, X. Li, M. Stappers, M. Le Gallo, X. Fu, A. Lukashchuk, A. S. Raja, J. Liu, C. D. Wright, A. Sebastian, T. J. Kippenberg, W. H. P. Pernice, and H. Bhaskaran, ``Parallel convolutional processing using an integrated photonic tensor core,'' Nature \textbf{589}, 52--58 (2021).


\bibitem{jiangNRM20} J. Jiang, M. Chen, and J. A. Fan, ``Deep neural networks for  the evaluation and design of photonic devices,'' Nature Reviews Materials \textbf{6}, 679--700 (2021).

\bibitem{hughes18} T. W. Hughes, M. Minkov, Y. Shi, and S. Fan, ``Training of photonic neural networks through in situ back-propagation and gradient measurement,'' Optica \textbf{5}, 864--871 (2018).

\bibitem{zhouPR20} T. Zhou, L. Fang, T. Yan, J. Wu, Y. Li, J. Fan, H. Wu, X. Lin, and Q. Dai, ``In situ optical backpropagation training of diffractive optical neural networks,'' Photonics Research \textbf{8}, 940--953 (2020).

\bibitem{stenbrecher20} G. R. Steinbrecher, J. P. Olson, D. Englund, and J. Carolan, ``Quantum optical neural networks,'' npj Quantum Information \textbf{5}, 60 (2019).

\bibitem{J. T. Connor14}J. T. Connor, R. D. Martin, and L. E. Atlas, ``Recurrent neural networks and robust time series prediction,'' IEEE Transactions on Neural Networks and Learning Systems \textbf{5}, 240--254 (1994).

\bibitem{G. Dorffner13} G. Dorffner, ``Neural networks for time series processing,''  Neural Networks World \textbf{6}, 447--468 (1996).

\bibitem{M. Husken12}M. H\"usken, and P. Stagge, ``Recurrent neural networks for time series classification,'' Neurocomputing \textbf{50}, 223--235 (2003).

\bibitem{K. Yao11} K. Yao, G. Zweig, M.-Y. Hwang, Y. Shi, and D. Yu, ``Recurrent neural networks for understanding,'' In Proceedings of Interspeech, 2524--2528 (2013).

\bibitem{I. Goodfellow 17} I. Goodfellow, Y. Bengio, and A. Courville, ``Deep Learning,'' The MIT Press \textbf{1}, 326--366 (2016).

\bibitem{by Hugeyyy}T. W. Hughes, I. A. D. Williamson, M. Minkov, and S. Fan, ``Wave physics as an analog recurrent neural network," Science Advances \textbf{5}, eaay6946 (2019).

\bibitem{yuanoptica18} L. Yuan, Q. Lin, M. Xiao, and S. Fan, ``Synthetic dimension in photonics,'' Optica \textbf{5}, 1396--1405 (2018).

\bibitem{ozawaNRP19} T. Ozawa, and H. M. Price, ``Topological quantum matter in synthetic dimensions,'' Nature Reviews Physics \textbf{1}, 349--357 (2019).

\bibitem{Topological photonics} E. Lustig, and M. Segev, ``Topological photonics in synthetic dimensions,'' Advances in Optics and Photonics \textbf{13}, 426--461 (2021).

\bibitem{Chinese Optics Letters} H. Liu, Z. Yan, M. Xiao, and S. Zhu, ``Recent Progress in Synthetic Dimension in Topological Photonics,'' Chinese Optics Letters \textbf{41}, 0123002 (2021).

\bibitem{APL photonics 6} L. Yuan, A. Dutt, and S. Fan, ``Synthetic frequency dimensions in dynamically modulated ring resonators,'' APL Photonics \textbf{6}, 071102 (2021).

\bibitem{pankov19} A. V. Pankov, O. S. Sidelnikov, I. D. Vatnik, A. A. Sukhorukov, and D. V. Churkin, ``Deep learning with synthetic photonic lattices for equalization in optical transmission systems,'' Proc. SPIE 11192, Real-time Photonics Measurements, Data Management, and Processing IV, 111920N (2019).


\bibitem{buddhiraju20} S. Buddhiraju, A. Dutt, M. Minkov, I. A. D. Williamson, and S. Fan, ``Arbitrary linear transformations for photons in the frequency synthetic dimension,'' Nature Communications \textbf{12}, 2401 (2021).

\bibitem{linArxiv20} Z. Lin, S. Sun, J. Azana, W. Li, N. Zhu, and M. Li, ``Temporal optical neurons for serial deep learning,'' arXiv:2009.03213 (2020).

\bibitem{regensburger11} A. Regensburger, C. Bersch, B. Hinrichs, G. Onishchukov, A. Schreiber, C. Silberhorn, and U. Peschel, ``Photon propagation in a discrete fiber network: an interplay of coherence and losses,'' Physical Review Letters \textbf{107}, 233902 (2011).

\bibitem{regensburger12} A. Regensburger, C. Bersch, M.-A. Miri, G. Onishchukov, D. N. Christodoulides, and U. Peschel, ``Parity-time synthetic photonic lattices,'' Nature \textbf{488}, 167--171 (2012).

\bibitem{wimmer13} M. Wimmer, A. Regensburger, C. Bersch, M.-A. Miri, S. Batz, G. Onishchukov, D. N. Christodoulides, and U. Peschel, ``Optical diametric drive acceleration through action-reaction symmetry breaking,'' Nature Physics \textbf{9}, 780--784 (2013).

\bibitem{marandi14} A. Marandi, Z. Wang, K. Takata, R. L. Byer, and Y. Yamamoto, ``Network of time-multiplexed optical parametric oscillators as a coherent Ising machining,'' Nature Photonics \textbf{8}, 937--942 (2014).

\bibitem{wimmer17} M. Wimmer, H. M. Price, I. Carusotto, and U. Peschel, ``Experimental measurement of the Berry curvature from anomalous transport,'' Nature Physics \textbf{13}, 545--550 (2017).

\bibitem{chenPRL18} C. Chen, X. Ding, J. Qin, Y. He, Y.-H. Luo, M.-C. Chen, C. Liu, X.-L. Wang, W.-J. Zhang, H. Li, L.-X. You, Z. Wang, D.-W. Wang, B. C. Sanders, C.-Y. Lu, and J.-W. Pan, ``Observation of topologically protected edge states in a photonic two-dimensional quantum walk,'' Physical Review Letters \textbf{121}, 100502 (2018).

\bibitem{23newlarger17} L. Larger, A. Bayl\'{o}n-Fuentes, R. Martinenghi, V. S. Udaltsov, Y. K. Chembo, and M. Jacquot, ''High-speed photonic reservoir computing using a time-delay-based architecture: million words per second classification,'' Phys Review X \textbf{7}, 011015 (2017).

\bibitem{arxiv9-8Aus}A. V. Pankov, I. D. Vatnik, A. A. Sukhorukov, ``Optical neural network based on synthetic nonlinear photonic lattices," Physical Review Applied \textbf{17}, 024011 (2022).


\bibitem{zzzmnist} Y. Lecun, and L. Botto, ``Gradient-based learning applied to document recognition," Proceedings of the IEEE \textbf{86}, 2278--2324 (2020).

\bibitem{leefmans} C. Leefmans, A. Dutt, J. Williams, L. Yuan, M. Parto, F. Nori, S. Fan, and A. Marandi, ``Topological dissipation in a time-multiplexed photonic resonator network,'' Nature Physics \textbf{18}, 442 (2022).

\bibitem{baoNR11} Q. Bao, H. Zhang, Z. Ni, Y. Wang, L. Polavarapu, Z. Shen, Q.-H Xu, D. Tang, and K. P. Loh,  ``Monolayer graphene as a saturable absorber in a mode-locked laser,'' Nano Research \textbf{4}, 297--307 (2011).

\bibitem{chengIEEE14} Z. Cheng, H. K. Tsang, X. Wang, K. Xu, and J.-B. Xu, ``In-plane optical absorption and free carrier absorption in graphene-on-silicon waveguides,'' IEEE Journal of Selected Topics in Quantum Electronics \textbf{20}, 43--48 (2014).

\bibitem{nonlinearnew3} Q. Xie, H. Zhang, and C. Shu, ``Programmable Schemes on Temporal Waveform Processing of Optical Pulse Trains,'' Journal of Lightwave Technology \textbf{38}, 339--345 (2020).

\bibitem{bengio09} Y. Bengio, ``Learning deep architectures for AI,'' Foundations and Trends in Machine Learning \textbf{2}, 1--127 (2009).

\bibitem{pooling} D. Scherer, A. M{\"u}ller, and S. Behnke, ``Evaluation of Pooling Operations in Convolutional Architectures for Object Recognition," Proceedings of 20th International Conference on Artificial Neural Networks \textbf{6354} LNCS (PART 3), 92--101 (2010).

\bibitem{newRaudys} S. Raudys, ``Evolution and generalization of a single neurone: I. Single-layer perceptron as seven statistical classifiers,'' Neural Networks \textbf{11}, 283--296 (1998).

\bibitem{newLehtokangas} M. Lehtokangas and J. Saarinen, ``Weight initialization with reference patterns,'' Neurocomputing \textbf{20}, 265--278 (1998).

\bibitem{newSebastiani} F. Sebastiani, ``Machine Learning in Automated Text Categorization,'' ACM Computing Surveys \textbf{34}, 1--47 (2002).

\bibitem{newSaleem} N. Saleem and M. I. Khattak, ``Deep neural networks based binary classification for single channel speaker independent multi-talker speech separation,'' Applied Acoustics \textbf{167}, 107385 (2020).



\bibitem{previous onn2} D. Psaltis, D. Brady, and K. Wagner, ``Adaptive optical networks using photorefractive crystals,'' Applied Optics \textbf{27}, 1752--1759 (1988).





\bibitem{nonlinearnew1} S. Tainta, M. J. Erro, W. Amaya, M. J. Garde, S. Sales, and M. A. Muriel, ``Periodic time-domain modulation for the electrically tunable control of optical pulse
train envelope and repetition rate multiplication,'' IEEE Journal
of Selected Topics in Quantum Electronics \textbf{18}, 377--383
(2012).

\bibitem{nonlinearnew2} A. Malacarne and J. Aza\~{n}a, ``Discretely tunable comb spacing of a frequency comb by multilevel phase modulation of a periodic pulse train,'' Optics Express \textbf{21}, 4139--4144 (2013).

\bibitem{IEEE-William} I. A. D. Willianmson, T. W. Hughes, M. Minkov, B. Bartlett, S. Pai, and S. Fan, ``Reprogrammable electro-optic nonlinear activation functions for optical neural networks," IEEE Journal of Selected Topics in Quantum Electronics \textbf{26}, 7700412 (2020).

\bibitem{chensegev} Z. Chen and M. Segev, ``Highlighting photonics: looking into the next decade,'' eLight \textbf{1}, 2--12 (2021).


\bibitem{newSierra} E. Duran-Sierra, S. Cheng, R. Cuenca, B. Ahmed, J. Ji, V. V. Yakovlev, M. Martinez, M. Al-Khalil, H. Al-Enazi, Y.S. L. Cheng, J. Wright, C. Busso, J. A Jo, ``Machine-learning assisted discrimination of precancerous and cancerous from healthy oral tissue based on multispectral autofluorescence lifetime imaging endoscopy,'' Cancers \textbf{13}, 4751 (2021).

\bibitem{newShirshin} E. A. Shirshin, A. V. Gayerm, E. E. Nikonova, M. M. Lukina, B. P. Yakimov, G. S. Budylin, V. V. Dudenkova, N. I. Ignatova, D. V. Komarov, E. V. Zagaynova, V. V. Yakovlev, W. Becker, V. I. Shcheslavskiy, M. Shirmanova, and M. O. Scully, ``Label-free sensing of cells with fluorescence lifetime imaging: the quest for metabolic heterogeneity,'' Proceedings National Academy of Sciences USA \textbf{119}, e2118241119 (2022).

\end{thebibliography}
\end{document}